# Resolving the Interstellar Medium at the Peak of Cosmic Star Formation

Gabriela Calistro Rivera [1]
Jacqueline Hodge [1]

For the ALESS consortium

[1] Leiden Observatory, the Netherlands

**The interstellar medium feeds both the formation of stars and the growth of black holes, making it a key ingredient in the evolution of galaxies. With the advent of the Atacama Large Millimeter/submillimeter Array (ALMA), we can now probe the interstellar medium within high-redshift galaxies in increasingly exquisite detail. Our recent ALMA observations map the molecular gas and dust continuum emission in submillimetre-selected galaxies on 1–5 kpc scales, revealing significant differences in how the gas, dust continuum, and existing stellar emission are distributed within the galaxies. This study demonstrates the power of ALMA to shed new light on the structure and kinematics of the interstellar medium in the early Universe, suggesting that the interpretation of such observations is more complex than typically assumed.**

## The interstellar medium in the early Universe

The growth of a galaxy, through the formation of new stars, is deeply connected to its interstellar medium. Characterising the interstellar medium within galaxies thus sheds light on the physical conditions around the process of star formation. The submillimetre and far-infrared regimes play a prominent role in this characterisation, since both the line and continuum emission are largely related to the star formation process in galaxies. Indeed, the emission lines in this regime trace the cold molecular gas phase most closely related to the feeding mechanism of star formation (for example, Narayanan et al. 2011). At the same time, the peak of the dust continuum emission in galaxies is considered to be one of the most reliable tracers of recent star formation episodes, since it is the optical and ultraviolet emission from newly-born stars that heat the dust to produce the far-infrared continuum emission.

Among the molecular gas emission lines, carbon monoxide (CO) is the most strongly emitting molecule and thus the most commonly used molecular gas tracer. However, it is only the second most abundant molecule in the interstellar medium after molecular hydrogen, $H_2$, which dominates by mass. Since $H_2$ is not easily observable, most studies use CO emission as a tracer of the molecular mass. This, however, requires a CO-to-$H_2$ conversion factor ($\alpha_{CO}$). Numerous observational and theoretical efforts suggest that this factor has strong dependencies on galaxy properties such as gas density, temperature and metallicity, although these relationships are not well understood, particularly in the high-redshift Universe (for example, Bolatto et al. 2013). Although this conversion factor plays a key role in molecular gas measurements, it represents one of the biggest uncertainties in high-redshift molecular gas studies.

Since detecting the CO emission from distant galaxies can still require long exposure times, another way to indirectly probe the cold interstellar medium is through observations of the cold Rayleigh-Jeans tail of the dust continuum emission. Efforts to study the evolution of the cold gas content of galaxies have increasingly relied on such unresolved dust continuum measurements to construct statistically significant samples (see, for example, Scoville et al. 2015). These studies carry their own assumptions, including that the ratio of gas to dust is constant, and that these components are well-mixed and co-located within galaxies.

To test the validity of these assumptions, detailed observations of both gas and dust continuum emission within galaxies at similar spatial resolutions are required. Galaxies selected on the basis of their bright emission in the submillimetre continuum (known as submillimetre galaxies, or SMGs) are excellent laboratories in which to carry out these observations due to their large molecular reservoirs and bright dust continuum emission. In addition, SMGs contribute around 20% of the total star formation rate density at $z \sim 2$–3 (Swinbank et al., 2014), making them an important population to study in order to characterise the peak epoch of galaxy assembly.

## A spatially resolved study of the molecular gas and dust in submillimetre galaxies

The long baselines and large collecting area provided by ALMA make deep, high-resolution observations of multiple tracers of the interstellar medium increasingly possible for distant galaxies and, most crucially, in a fraction of the time required by other facilities. We therefore exploited the capabilities of ALMA to resolve the molecular gas emission — through CO — and dust continuum emission in a sample of bright submillimetre galaxies at redshift $z \sim 2.5$.

The SMGs in our study are part of the ALMA LABOCA ECDFS Submillimeter Survey (ALESS; Hodge et al., 2013). This is an ALMA Cycle 0 study of ~ 100 luminous and ultra-luminous infrared galaxies in the Extended Chandra Deep Field (ECDFS) originally selected based on the Atacama Pathfinder Experiment (APEX) LArge APEX BOlometer Camera (LABOCA). The galaxies in the ALESS survey constitute the most extensively studied submillimetre galaxies to date in terms of multi-wavelength and spectroscopic follow-up observations, including accurate spectroscopic redshifts, radio to X-ray photometric characterisation, and now high-resolution ALMA follow-up (Hodge et al., 2016; Chen et al., 2017; Calistro Rivera et al., 2018).

In Calistro Rivera et al. (2018), we present the latest high-resolution (subarcsecond) molecular gas observations of the CO($J$ = 3–2) transition in four ALESS SMGs: ALESS 49.1, 57.1, 67.1 and 122.1 (Project ID 2013.1.00470; PI: Hodge). With only 30 minutes of exposure time per source with ALMA's Band 3, we detect and resolve the CO(3–2) emission in these high-redshift galaxies (Figure 1). The measured positions are coincident with the ALMA dust continuum emission, and the measured frequencies confirm the expected spectroscopic redshifts ($z \sim 2$–3). Figure 1 shows the CO(3–2) emission overlaid on the stellar emission as traced by the Wide Field Camera 3 (WFC3) and/or Advanced Camera for Surveys (ACS) imaging from the Hubble Space Telescope (HST). With the exception of ALESS 122.1, the CO gas overlaps the existing stellar distributions.

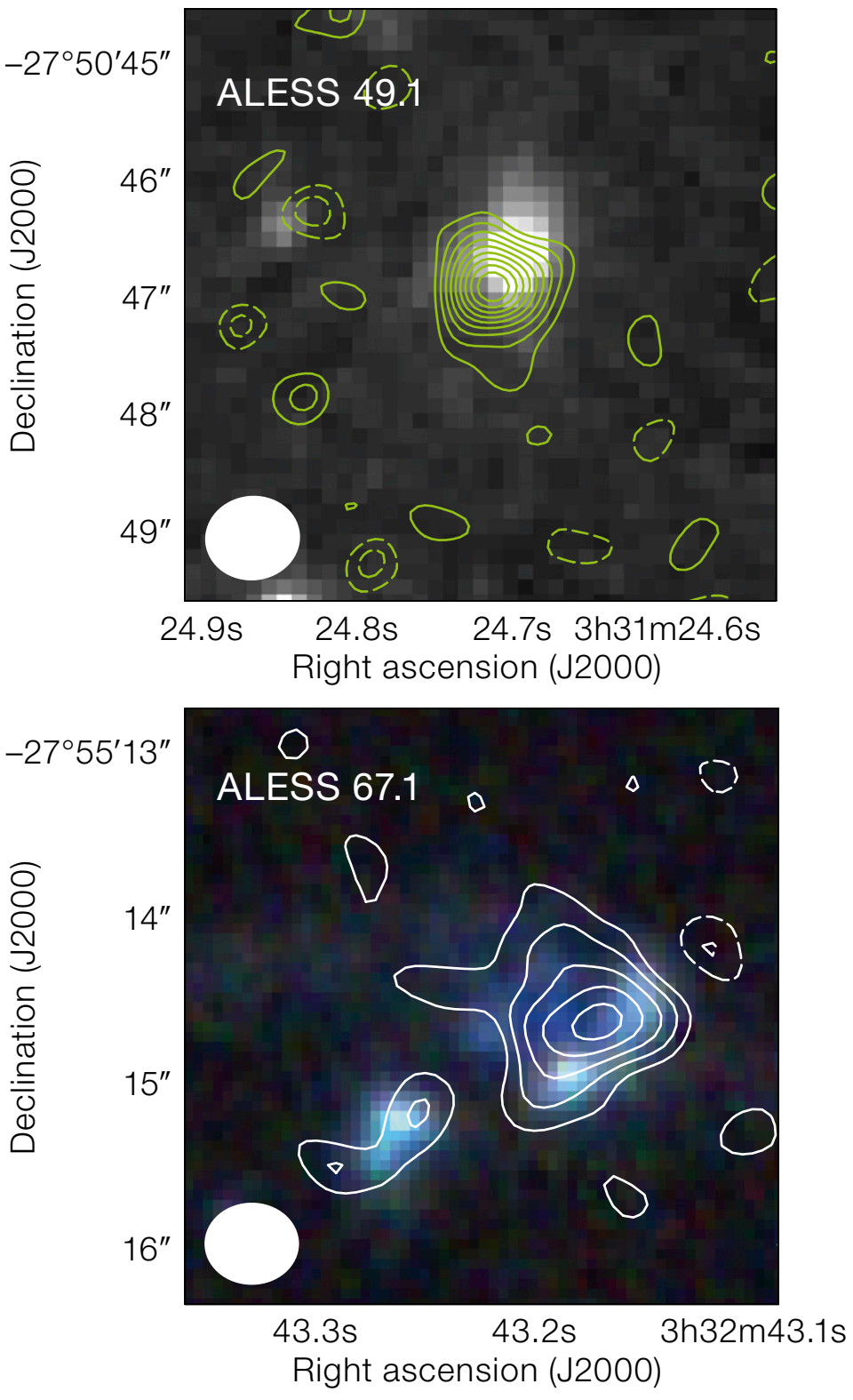

Figure 1. The molecular gas emission, as traced by the velocity-averaged CO(3–2) emission (Calistro Rivera et al., 2018), is shown as contours overlaid on the stellar emission, traced by the Hubble Space Telescope imaging. We see that our 30-minute-per-source ALMA observations can strongly detect and resolve the CO emission in these four SMGs on scales of around four kiloparsecs (beam sizes are shown in the lower-left corners), showing a diversity of morphologies in both CO and stellar emission. The accurate astrometry, as calibrated using Gaia data, reveals offsets among the gas and stellar emission distributions.

High-resolution 870-micron continuum observations for 16 ALESS galaxies were previously presented by Hodge et al. (2016). Figure 2 shows that the dust continuum emission in these galaxies is distributed over scales of only a few kiloparsecs and appears to have a smooth and disk-like morphology at the sensitivity and resolution of the observations. In Chen et al. (2017), we presented a detailed comparison of the dust continuum and CO for one source. Here, we use all of the high-resolution gas and dust continuum data now available for the ALESS sources to learn about the resolved properties of the interstellar medium in the early Universe.

## A census of gas masses using galaxy kinematics

The CO(3–2) data obtained in these observations have the potential to reveal the total mass of molecular gas in these SMGs. However, the derivation of total masses from CO data depends on several unknown factors, which can now be inferred from our observations.

The total mass of a galaxy within a given radius can be estimated from the dynamics of the contained matter according to the virial theorem. These kinematic properties are reflected in the CO line emission and can be extracted by applying kinematic modelling to the velocity field observed in the line (or in cases where the resolution or signal-to-noise ratio is not enough, they can be approximated simply by the observed line-widths). In Figure 3, we use a model of a rotating exponential disk galaxy (Galpak3D; Bouché et al., 2016) in order to fit the CO(3–2) emission of the source of highest signal-to-noise ratio in our sample, ALESS 122.1. We infer the relevant kinematic parameters that produce the best fit, such as the inclination angle and maximum rotational velocity of the gas. These parameters allow us to estimate the total dynamical mass of the galaxy.

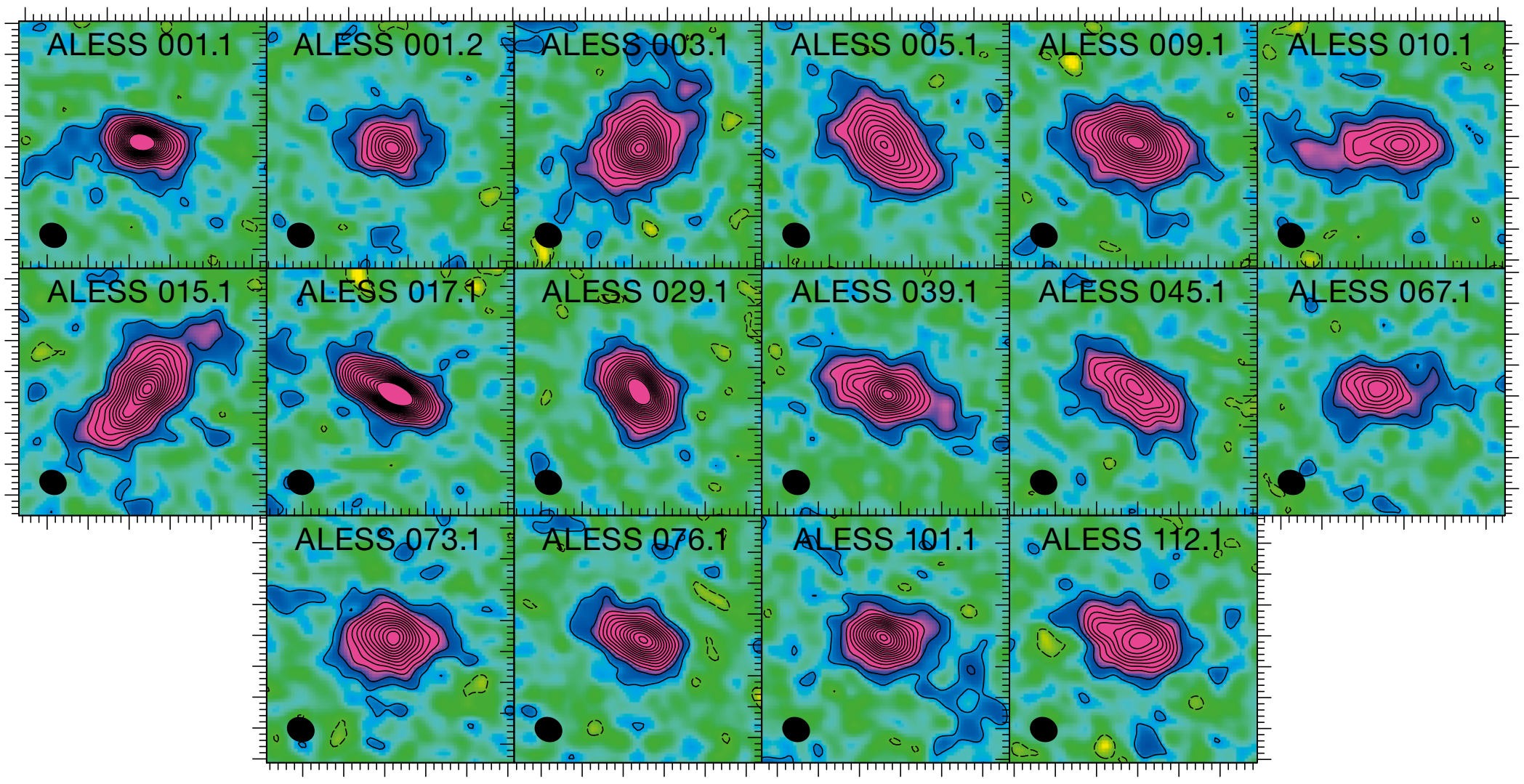

Figure 2. The 870-μm dust continuum emission at around one kiloparsec resolution in 16 submillimetre galaxies from the ALESS survey (Hodge et al., 2016). The dust continuum emission is distributed over scales of only a few kiloparsecs and appears to have a smooth and disc-like morphology at the sensitivity and resolution observed.

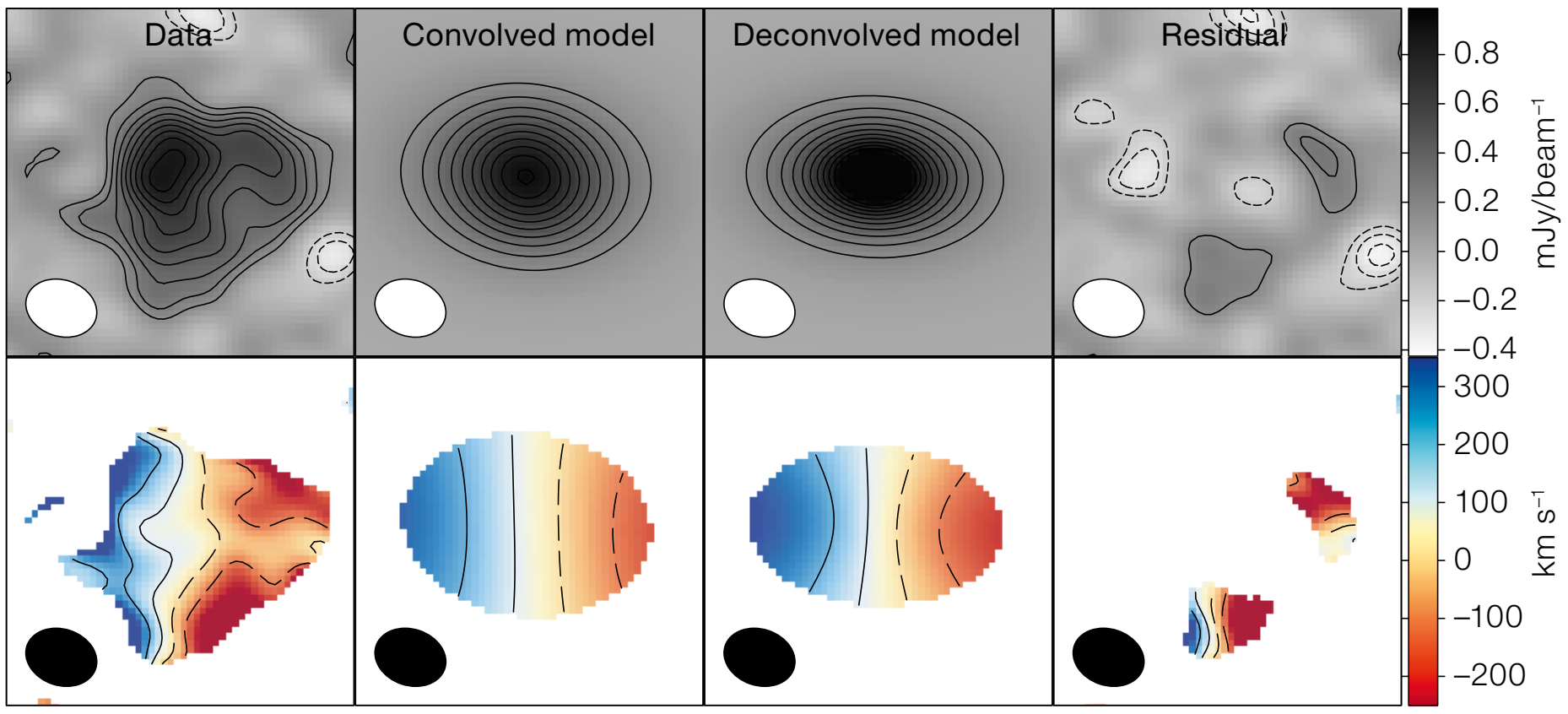

Figure 3. Kinematic modelling of the CO emission in ALESS 122.1 using the code Galpak3D. The top row shows the integrated intensity maps. The bottom row shows the intensity-weighted velocity maps. The field of view of each image is ~ 1 square arcminute. The columns correspond to the observed data, the best-fit model convolved with the beam, the deconvolved best-fit model and the residual. The cool molecular gas emission in this source, as traced by the CO(3–2) emission, can be well described by a rotating disk. Given the clumpy stellar emission in the source, this result suggests that an ordered rotating molecular gas disk could have been quickly reformed after a merger event.

The mass of a galaxy is essentially composed of dark matter, stars and the gas that constitutes its interstellar medium. A census of the various components contributing to the mass of a galaxy can shed light on unknown parameters needed in molecular gas studies, such as the CO-to-$H_2$ conversion factor, $\alpha_{CO}$, mentioned above. Through our estimated total dynamical masses — as well as further knowledge and assumptions about the dark matter and stellar mass components — we can then estimate the remaining mass which is in the form of gas. Since the total gas mass is predominantly molecular hydrogen at these redshifts, it provides a fair estimate of the hydrogen mass content in the ISM of the galaxy. Relating this estimate of the total molecular hydrogen mass to the measured CO luminosity, we can finally infer the conversion factor $\alpha_{CO}$.

In applying this method in our study, we put particular emphasis on recovering robust uncertainties in the estimation of $\alpha_{CO}$. These uncertainties arise from assumptions made about the dark matter fraction and stellar mass component, in particular. We use a Bayesian method in order to sample the probability density functions of all of the unknown parameters contributing to these uncertainties, such as the mass-to-light ratio of the stellar component and the unknown dark matter fraction. Taking advantage of our high-resolution imaging, we are able to recover a mean value of $\alpha_{CO} = 0.9 \pm 0.6$ ($M_\odot$/[K km s$^{-1}$ pc$^2$]$^{-1}$). This value is consistent with the value generally assumed for luminous and ultra-luminous infrared galaxies in the local Universe, but is considerably smaller than the Milky Way value ($\alpha_{CO}$ ~ 4). We also investigate the covariance between $\alpha_{CO}$ and the stellar mass-to-light ratio, showing that they are strongly correlated. Although we are limited here by a small sample of four SMGs, we suggest that this method holds great potential for robustly recovering these unknown parameters when better statistics are available.

## Offset distributions of stars, gas and dust

The availability of high-resolution, subarcsecond imaging at different wavelengths, such as the optical/near-infrared imaging of the Hubble Space Telescope and the far-infrared dust continuum emission mapped by ALMA, allows us to make a detailed image of the different physical components within a high-redshift galaxy. In Figure 4, we explore the case of the offset distribution of the CO gas (green region) relative to the stellar emission (blue region) in ALESS 122.1, now including the dust continuum emission (red region) corresponding to the rest-frame 1 mm emission. The emission from the cold dust appears to be colocated with the gas, as expected since both trace the interstellar medium.

However, the emission from the dust does not overlap with the existing stellar emission, suggesting that they may not be produced in the same regions. This finding has important implications for methods commonly used at high redshift to fit panchromatic spectral energy distributions — particularly energy balance techniques that attempt to self-consistently model the observed optical and infrared emission by coupling the dust and stellar emission. The increasing number of multi-wavelength, high-resolution observations at high redshift with similar findings are indeed challenging this picture, especially for extreme sources such as SMGs (Hodge et al., 2016).

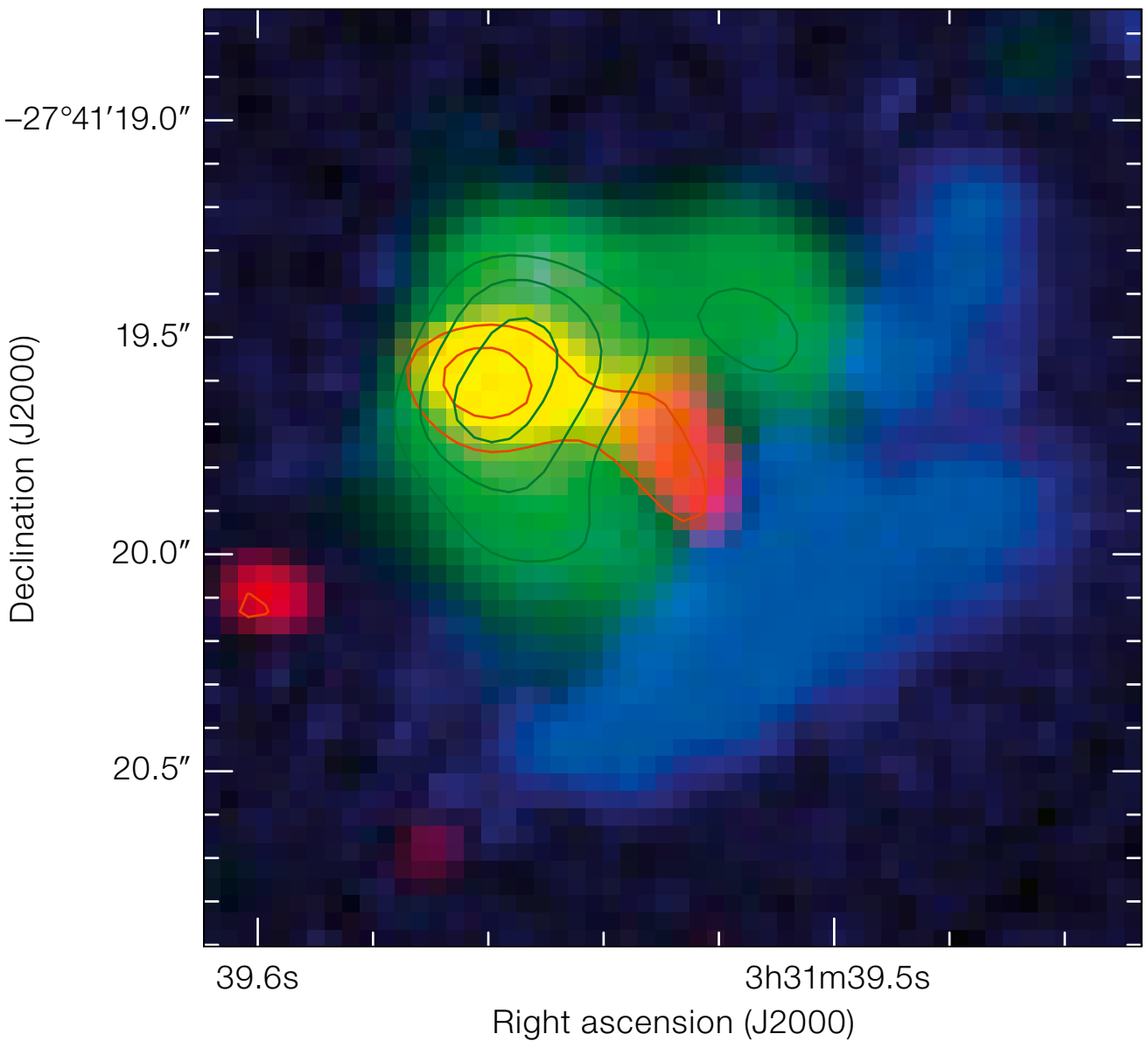

Figure 4. Image of the cool molecular gas emission (green), dust continuum emission (red) and stellar emission (blue) components in the submillimetre galaxy ALESS 122.1. The physical components in this galaxy appear clearly misaligned. This galaxy is one example of a large number of similar cases within high-resolution submillimetre surveys of extreme sources such as SMGs (Hodge et al., 2016). These observations may have important implications for energy-balance assumptions between the integrated dust and stellar emission of extreme sources.

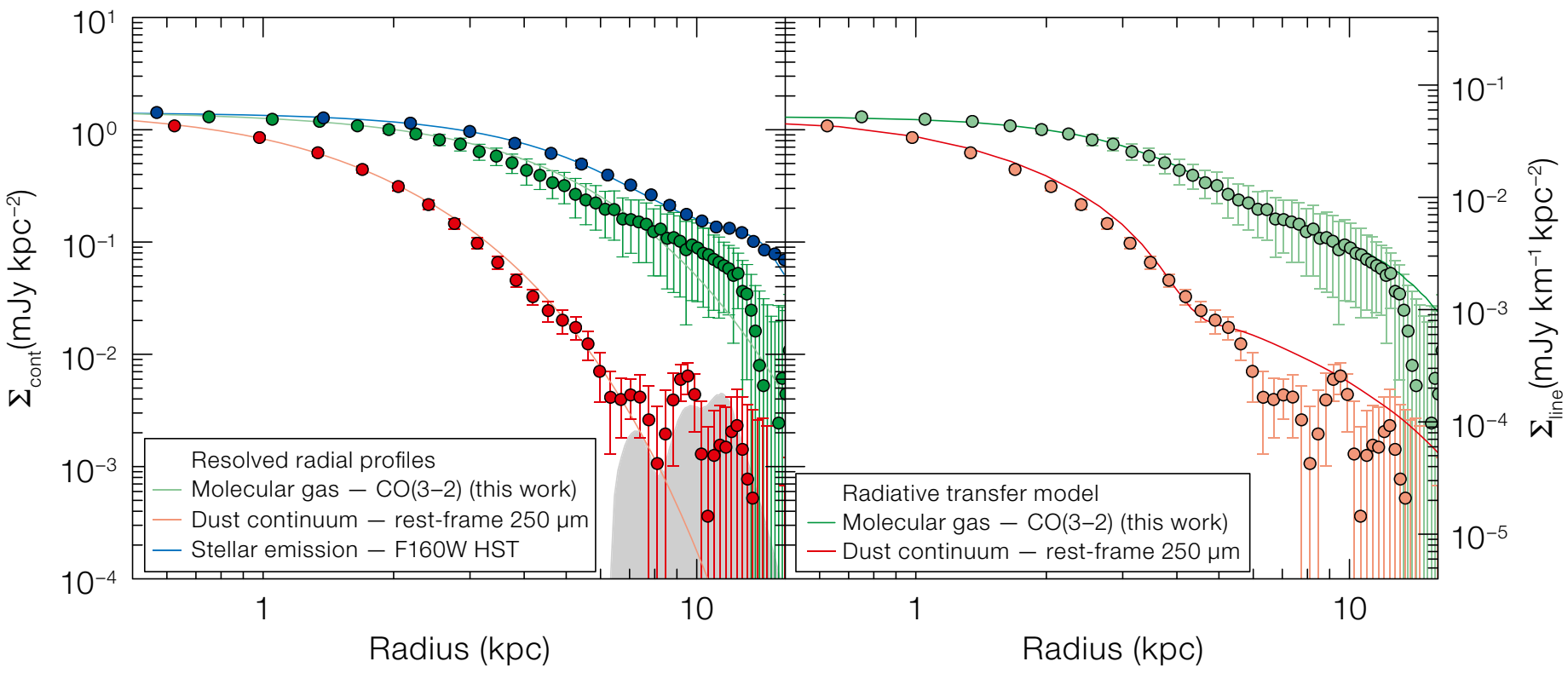

Figure 5. Stacked radial profiles of the gas (green), dust (red) and stellar (blue) emission in ALESS submillimetre galaxies, showing the average intensity as a function of radius. The solid lines in the left panel show the best fit exponential profiles, which have been convolved with Gaussian functions of the size of the ALMA beam to account for the effect of the different resolutions. We note that the data contaminated by the beam side lobes (grey shaded area) have been removed for the fit. This plot shows that the dust continuum emission is more than twice as compact as the stellar and molecular gas emission, which have similar extents. In the right panel, we use radiative transfer to consistently model the gas and dust phase of the interstellar medium in these sources (Weiss et al., 2007). The model shows that the apparent difference between the dust and gas emission sizes does not necessarily imply different intrinsic physical distributions, but rather can arise from temperature and optical depth gradients alone.

## A statistical study of the distribution of the gas and dust emission

Although not all SMGs show an offset between their dust and stellar distributions, a growing number of studies find a contrast between the apparent sizes of the compact dust and extended gas and stellar components, suggesting that this difference in physical scale is a general feature. In order to investigate this observation in a statistical manner, in Figure 5 we apply a stacking analysis to all ALESS submillimetre galaxies with high-resolution dust, gas and stellar emission imaging. To produce the average gas radial profile (green points), we stack the profiles of the four ALESS galaxies in Figure 1 presented by Calistro Rivera et al. (2018). The average radial profiles of the dust (red points) and stellar emission (blue points) are produced through stacking 16 ALESS galaxies presented in Figure 2 and by Hodge et al. (2016). Through the stacking method, we not only investigate the average properties of the population, but also achieve a higher sensitivity since the signal-to-noise ratio improves in the final stacked image.

Our statistical approach reveals that the (observed-frame) 870 μm dust continuum extends over less than half of the gas emission, while the gas and the stellar emission distributions appear to have similar spatial extents. This finding raises several questions regarding the physical nature and relationship of the emitting components. In particular, is the compact dust distribution an observational effect due to the lack of sensitivity? Despite the higher sensitivity achieved through the stacking, we do not recover a significant low-surface-brightness component to the dust emission. Is the dust continuum then physically present only in a central compact region, while the gas is more spread out? This would mean that gas and dust are not well-mixed in the interstellar medium as is commonly assumed. To answer this last question, we apply radiative transfer modelling to our observations in order to recover the physics that may produce these different distributions.

We use an updated version of the radiative transfer model of Weiß et al. (2007) to simulate the stacked radial profiles of the gas and dust as presented in the right panel of Figure 5. In this method, we consistently relate the gas and the dust emission through radiative transfer by assuming that gas and dust are well mixed and equally distributed throughout the galaxy. As initial conditions, we apply a radially decreasing temperature and optical depth gradient to the system, as commonly expected in centrally located starbursts. The right panel of Figure 5 shows the best-fit model to the data, demonstrating that the different apparent extents of dust and gas can be well reproduced by the proposed model.

These results indicate that the apparent difference in size observed between the compact dust continuum and extended gas emission does not necessarily imply different physical distributions for these components. Rather, this apparent size difference can be a consequence of a combination of temperature and optical depth gradients, which can be extreme in galaxies such as SMGs, but also non-negligible in normal star-forming galaxies. Such a scenario has important implications for methods that adopt dust continuum sizes as an approximation to molecular gas sizes, especially for the calculation of dynamical masses. These results demonstrate the potential of high-resolution campaigns with ALMA to reveal observational challenges and solve open questions in the study of the interstellar medium in the early Universe.

### Acknowledgements

We thank Leonard Burtscher for useful comments that improved the article. Jacqueline Hodge acknowledges support of the VIDI research programme with project number 639.042.611, which is (partly) financed by the Netherlands Organisation for Scientific Research (NWO).